\documentclass[aps,prb,twocolumn,floatfix]{revtex4}

\usepackage{amssymb}
\usepackage{graphicx}

\def\be{\begin{equation}}
\def\ee{\end{equation}}
\def\bea{\begin{eqnarray}}
\def\eea{\end{eqnarray}}

\begin{document}
\title{Non-equilibrium phonon transport in surface-roughness dominated nanowires}
\author{S. Abhinav and K. A. Muttalib}
\affiliation
{Department of Physics, University of Florida, Gainesville, FL 32611-8440, USA}
\begin{abstract}

Experimental observation of highly reduced thermal conductivity in surface-roughness dominated silicon nanowires have generated renewed interest in low-dimensional thermoelectric devices. Using a previous work where the scattering of phonons from a rough surface is mapped to scattering from randomly situated localized phonons in the bulk of a smooth nanowire, we consider the thermal current across a nanowire for various strengths of surface disorder. We use non-equilibrium Green's function techniques that allow us to evaluate the thermal current beyond the linear response regime, for arbitrary cold and hot temperatures of the two semi-infinite connecting leads. We show how the surface-roughness affects the frequency dependence of the thermal current, eventually leading to a temperature dependent reduction of the net current at high temperatures. We use a universal disorder parameter to describe the surface-roughness as has been proposed, and show that the dependence of the net current on this  parameter provides a natural explanation for the experimentally observed differences between smooth vs rough surfaces. We argue that a systematic study of the thermal current for different values of the temperature difference between the two sides of a surface-roughness dominated nanowire for various strengths of disorder would help in our understanding of how best to optimize the thermoelectric efficiency.

\end{abstract}

\maketitle
 
\section{Introduction}

Efficiency of thermoelectric devices are very sensitive to the thermal transport properties of the device elements \cite{book,slack,takabatake}. In particular in low-dimensional systems  where Fermi-liquid approximations break down, a good charge conductor can also have low thermal conduction by electrons as required for a good thermoelectric device \cite{hicks,majumdar,datta1,dresselhaus,reddy,abbout,wu,reviews,snyder}. However, heat carried by phonons can be much larger than heat carried by electrons and therefore can seriously reduce the overall efficiency of such a device. For example, it has been argued that silicon nanowires with appropriately tuned gate voltages can be designed to have very high efficiency as well as large power output \cite{hmn,mh}, but thermal conduction by phonons is an independent issue and needs to be studied independently. Fortunately, it has been shown experimentally that  if the silicon wires have strong surface disorder, thermal conductivity can be reduced well below the Casimir limit \cite{hochbaum,huang,lim}. Since similar surface disorder is expected to affect the electrical conductivity only weakly when the electron mean free path is smaller than the diameter \cite{tesanovich}, this highlights the importance of surface disorder as opposed to bulk disorder in nanowires in the context of thermoelectricity.

The effect of surface disorder on phonon transport in nanosystems has been studied numerically using a variety of techniques \cite{MC1,MC2,MD1,MD2,MD3,MD4,WS1,WS2,WS3,WS4,WS5,WS6,martin,murphy,chen,sadhu} including Monte Carlo and molecular dynamics as well as models using wave-scattering formalism. Such techniques rely on careful modeling of realistic surface disorder and its effects on elastic waves inside the system. A very different approach was proposed in Ref.~[\onlinecite{ma}], based on a mathematical mapping of propagating phonons scattering from the rough surface of a nanowire to their scattering from randomly situated localized phonons in the bulk of a smooth nanowire. This allowed the construction of a simple analytically tractable model that can provide useful insights into the nature of the roughness and its effects on the thermal conductivity. This work was limited to zero temperature only and coupling of the propagating phonons to the localized phonons was treated within only the leading order in perturbation theory. In the present work we extend this approach to a calculation of the thermal current in the presence of finite temperature difference $\Delta T\equiv T_L-T_R$ between the left (hot) and right (cold) lead temperatures $T_L$ and $T_R$, respectively; we use non-equilibrium Green's function techniques which allow us to go beyond the linear response regime.  In addition, we also evaluate the relevant non-equilibrium Green's functions self-consistently in order to extend the work beyond the weak coupling perturbative regime. Both extensions are important in the context of thermoelectricity in nanowire-based devices. Considering finite $\Delta T$ is important because it has been shown that the efficiency of a device can be increased significantly in the non-linear regime \cite{hmn,mh}, and also because in many practical applications (like harnessing the waste heat from a hot automobile engine at $T_{hot}\sim 450$ K in a cooler surrounding at $T_{cold}\sim 300$ K) a device must take advantage of the large temperature difference between the hot and cold leads. In addition, while the coupling of the propagating phonon to localized phonons might generally be weak for weakly disordered systems, a small thermal current needed for efficient devices requires large surface disorder, which in turn suggests the need to go beyond the weak coupling limit. 

In order to understand the qualitative features of the model unambiguously, we consider a simple model with only one localized phonon. This already captures an essential experimental observation that surface disorder leads to a frequency-dependent scattering \cite{lim}. On the basis of our understanding of the effects of a single localized phonon, we argue that a systematic measurement of the frequency-dependent thermal current  in such systems can provide important information about the distribution of such localized phonons. This, in turn, would allow more detailed theoretical predictions for phonon transport in these systems. Following Ref.~[\onlinecite{mm}], we take advantage of the fact that the effects of surface disorder, characterized by the mean surface corrugation height $h$, a roughness correlation length $l_c$ as well as the diameter $d$ of the wire, can be described by a single combination of the parameters. We show how the temperature difference $\Delta T$ between the leads affects the frequency-dependent current $J(\omega)$ as well as the net current $J_{net}$ for a given value of this universal disorder parameter. The non-linear thermal conductance  defined as $\kappa_{nl}\equiv J_{net}/\Delta T$ turns out to be a highly non-linear function of $\Delta T$, and this function changes with disorder. Thus we show that a linear response theory where the thermal conductance is treated as a constant depending only on the microscopic material parameters is clearly inadequate for large $\Delta T$. 

Although it is difficult to compare our results with existing experiments, they are qualitatively consistent with experiments on silicon nanowires with different strengths of surface disorder  \cite{hochbaum,huang,lim}.  In particular, it provides a natural explanation for the difference in thermal transport between smooth vs rough surfaces, e.g. the vapor-liquid-solid grown (VLS) \cite{huang} smooth nanowires  vs electrolessly etched (ELE) \cite{hochbaum} or electron beam lithography defined (EBL) \cite{huang} rough nanowires, in terms of the existence of and the strength of coupling to the localized phonons. We find that the effects of scattering from the localized phonons are more pronounced for larger temperature difference between the leads. We therefore propose that such effects have a better chance of being observed beyond the linear response regime.


\section{Theoretical model}

\begin{figure}
\begin{center}
\includegraphics[width=0.48\textwidth]{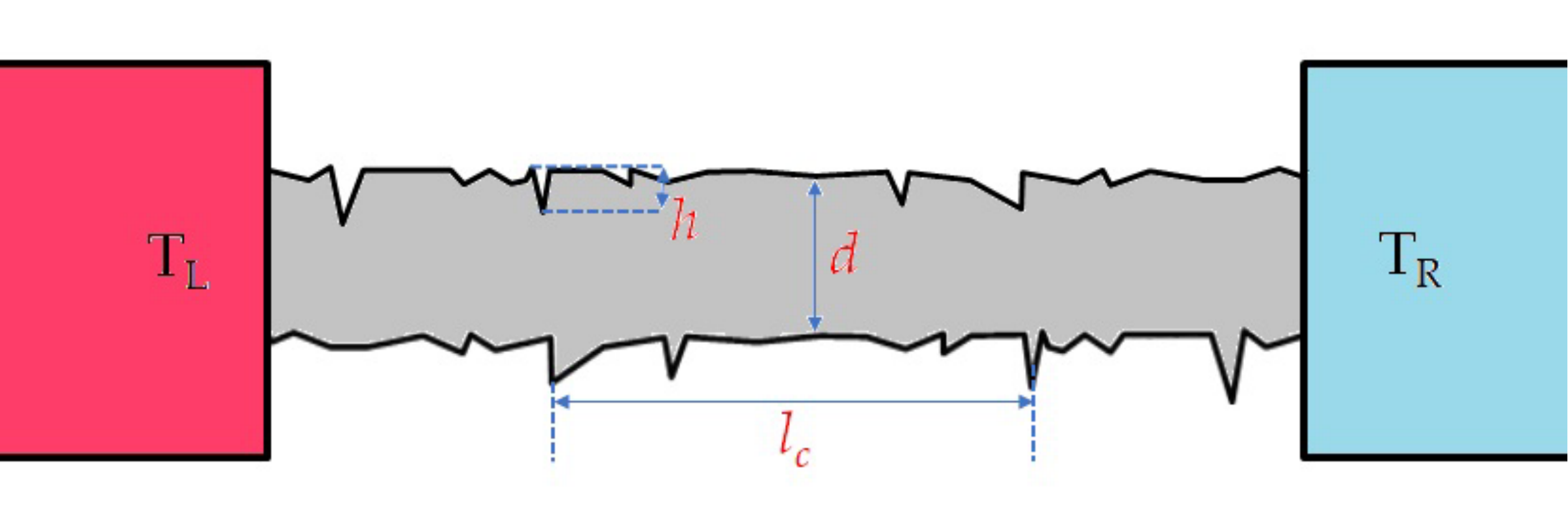}
\end{center}
\caption{
(Color online) A nanowire with surface roughness connected to leads at different temperatures $T_L$ (hot) and $T_R$ (cold). The roughness is characterized by the mean corrugation height $h$, the correlation length $l_c$ and the diameter $d$. }
\label{wire}
\end{figure}

We consider a nanowire with a fixed length much larger than the width, connected at the left and right ends to two semi-infinite leads, as shown in Figure \ref{wire}. 
While the wire along its length has surface disorder, the leads on two sides are assumed to be ideal, with no disorder. The left and right leads are kept at different temperatures $T_L$ and $T_R$ with $T_L > T_R$. We follow Ref.~[\onlinecite{ma}] to map this nanowire with a rough surface to a nanowire with smooth surface plus localized phonons situated at random positions in the bulk of the wire. 
Numerical simulations of surface-roughness dominated nanowires have indeed shown the existence of such localized phonons \cite{mm2019}. 
An averaging over the random positions leads to a self-energy for the propagating phonon (incident from the lead) of the form \cite{ma}
\be
\Sigma^{int}_p=N_{imp}\sum_{q,\nu}u_qu_{-q}d_q(\nu)D_{p-q}(\omega-\nu),
\label{sigma-int}
\ee
where $N_{imp}$ is the number of impurities over which the averaging is done, $u_qu_{-q}$ is related to the roughness power spectrum of the surface corrugation, and $d_q(\nu)$ and $D_{p-q}(\omega-\nu)$ are Green's functions corresponding to the localized and propagating phonons, respectively. From experiments, the following model for $u_qu_{-q}$ emerges \cite{ma,mm,lim}:
\be
u_qu_{-q}=W_0\frac{\Delta^2l_c}{1+q^2l_c^2}; \;\;\; \Delta=\frac{h}{d}
\ee
where the mean corrugation height $h$, the correlation length $l_c$ and diameter $d$ characterize the strength of surface disorder and $W_0$ determines the strength of the coupling of the propagating phonon with the localized phonon. The propagating phonon Green's function $D$ is obtained from the Dyson equation self-consistently, within the non-equilibrium framework \cite{rammer},
\be
D^{\lessgtr}=D^r\Sigma^{\lessgtr}D^a
\label{dyson}
\ee
where the superscripts $\lessgtr$ refer to the lesser and greater Green's function $D^{\lessgtr}$ and self energy $\Sigma^{\lessgtr}$, $r$ and $a$ refer to the retarded and advanced Green's functions,  and $\Sigma$ includes not only the interaction self energy $\Sigma^{int}$ given in (\ref{sigma-int}) but also the sum of self energies due to the two leads \cite{datta}.

The non-interacting lesser and greater propagating phonon Green's functions without coupling to the leads are of the form 
\bea  
D^{0,\lessgtr}_p(\omega)=-i2\pi\left[(1+ N_p)\delta(\omega\pm\omega_p)+N_p\delta(\omega\mp\omega_p) \right]
\label{D0}
\eea
where $N_p$ is the number of phonons at momentum $p$ and we assume an acoustic dispersion relation $\omega_p=vp$, $v$ being the sound velocity. We will use units where $v=1$.
Similarly the non-interacting localized phonon Green's function without coupling to the leads have a form similar to (\ref{D0})  with the propagating phonon frequency $\omega$ replaced by the localized phonon frequency $\Omega$.
We assume that the localized phonons couple weakly to the leads. This broadens the delta-function peaks providing a finite lifetime. Coupling to the leads can be incorporated in terms of the retarded Green's function $d^r(\omega)$ by including a self energy for the leads \cite{datta}
\bea
d^{\lessgtr}_q(\nu)&=&\frac{1}{2}\left[|d_q^r(\omega)|^2(\Sigma^{_0\lessgtr}_L(\omega)+ \Sigma^{_0\lessgtr}_R(\omega))\right. \cr
& + & \left.|d_q^r(-\omega)|^2(\Sigma^{_0\gtrless}_L(-\omega)+\Sigma^{_0\gtrless}_R(-\omega)\right] \cr
&=&-i\left[\frac{\Gamma_L(1+N_L(\omega))+\Gamma_R(1+N_R(\omega))}{(\nu\pm\Omega_q)^2+\Gamma^2} \right.\cr
&+& \left.\frac{\Gamma_LN_L(\omega)+\Gamma_RN_R(\omega)}{(\nu\mp\Omega_q)^2+\Gamma^2}\right].
\label{d1}
\eea
Here the subscripts $L$ and $R$ refer to the left and right leads, respectively. Thus $\Gamma_{L,R}$ are the widths of the localized phonon due to the coupling to the left and right leads, with $\Gamma_L+\Gamma_R=\Gamma$. We will assume symmetric coupling and use $\Gamma_L=\Gamma_R$. In (\ref{d1}) we have neglected the contribution to the real part of the self energy, because the frequency $\Omega_q$ will be used as a phenomenological parameter and any shift in its value corresponds simply to a different choice of the parameter. More importantly, the lead temperatures define the equilibrium Bose distribution functions:
\be
N_K(\omega)=\frac{1}{e^{\beta_K\omega}-1}, \;\;\; K=L,R.
\label{NK}
\ee
It takes into account the effects of the left and the right lead temperatures on the localized phonon Green's functions through the coupling to the leads. This is a significant simplification; it allows us to evaluate the non-equilibrium thermal current with the two leads at two different temperatures without having to define a non-equilibrium analog of the temperature inside the central wire.
  
Since the localized phonons arise from surface disorder, we expect a broad distribution of both $\Omega_q$  and $\Gamma$. Nevertheless for simplicity, we assume that there is only one localized phonon, of frequency $\Omega$. This is clearly a highly simplified model and one might worry that it would essentially limit the validity of the resulting conclusions. However, as we show below, choosing only one localized phonon allows us to investigate  its qualitative effects unambiguously. It captures the essential feature of the effect of surface roughness on thermal transport, namely a frequency dependent scattering rate \cite{lim}. The frequency dependent thermal current calculated later shows clearly what should be expected if more than one phonon or even a distribution of localized phonons is included. Such a distribution is not known at this point. On the other hand, if experimental measurements are available for the frequency dependent thermal current, our formulation will allow us to obtain insights into the distribution which, in turn, can be used to make further predictions.

We find that the correction to the localized phonon Green's function due to the interaction with the propagating phonons is negligible, so in our self-consistent calculation we use only the non-interacting (but coupled to the leads) form (\ref{d1}). 
The effect on the propagating phonon Green's function of adding leads is incorporated  using a lead self energy $\Sigma^{(L,R)}$ for left and right leads. The non-equilibrium current can be written in the form
\bea
J(\omega) =\omega\; {\rm Tr}\; \left[D^{>}_C(\Sigma^{<}_L-\Sigma^{<}_R)- D^{<}_C(\Sigma^{>}_L-\Sigma^{>}_R)\right]
\eea
where subscripts $L$ and $R$ refer to the self-energies due to the left and right leads, respectively, and $C$ refers to the central region, the nanowire. The Green's functions $D^{\lessgtr}_C$  are evaluated self-consistently from the non-equilibrium Dyson equation (\ref{dyson}). The leads are assumed to be ideal, semi-infinite and in equilibrium, with self energies due to the coupling to the central region given by
\bea
\Sigma^{>}_K &=&-i(1+N_K)\gamma^K, \;\;\; K=L,R \cr 
\Sigma^{<}_K &=&\frac{N_K}{1+N_K}\Sigma^{>}_K=-iN_K \gamma^K
\eea
where $N_{L,R}$ refer to the equilibrium Bose distributions (\ref{NK}) at the ideal leads kept in equilibrium at temperatures $T_L$ and $T_R$. 
We adopt a coupling to the lead $\gamma^K$  that takes into account the phonon density of states as well as any multi-phonon processes as proposed in Ref.~[\onlinecite{galperin}], given by
\be
\gamma^{K}(\omega)=\gamma_0^K\left(\frac{\omega}{\omega_c^K}\right)^2e^{2(1-\omega/\omega_c^K)}.
\label{multiphonon}
\ee
Here $\gamma_0^K$ is a constant of order $1$ and $\omega_c^K$ is a constant (less than the Debye frequency)
which is related to the multi-phonon process. Our qualitative results are not sensitive to the choices of these parameters.
Since the surface roughness is incorporated as localized phonons in the bulk, we can approximate our system as a quasi one-dimensional (1D) wire with 1D dispersion relation, avoiding a fully two-dimensional calculation that would be required for any direct phonon surface-roughness interaction model \cite{sadhu}. We will assume the diameter of the wire $d$ to be constant for our quasi 1D system, recognizing the fact that the effect of varying $d$ can not be included within this approximation. However as suggested in [\onlinecite{mm}], the effects of the diameter can be incorporated at least partly by including it in a universal disorder parameter, which we will discuss later.  

We use only the lowest order contribution to the interaction self energy neglecting all crossing diagrams. However the full Greens functions were evaluated self-consistently, which includes the leading exchange diagram (\ref{sigma-int}) to all orders. This allows us to go beyond the weak-coupling perturbative regime.


\section{Results}

Our main goal is to find the qualitative effects of surface roughness on the total  thermal current in the presence of a finite temperature difference between the two leads. To gain insights into the nature of the surface disorder, we first calculate the frequency dependence of the current, which gives the effect of a single localized phonon on the thermal current. As discussed above, we expect a distribution of such localized phonons which we can not model without experimental inputs. Nevertheless as we show below, since the effect of a single localized phonon is clearly identifiable,  we propose that comparing the single localized phonon with experiments should give us insights about the nature and properties of the distribution of localized phonons present in the system. 

Our temperatures are measured in units of Debye frequency $\omega_D$, with $\hbar=k_B=1$ such that $\omega_D=T_D=2\sqrt{2}$. We will choose $\gamma_0^K=1$ and $\omega_c^K=2$ in (\ref{multiphonon}). For simplicity we have used the same value of $\Omega$ for all our plots, namely $\Omega=1.5$. The width $\Gamma$ on the other hand depends on disorder, and we comment on our choice at the end of section A. Changing the values of these parameters  do not affect the qualitative aspects that we hope to understand. We will normalize our frequency dependent current $J(\omega)$ with the equilibrium thermal conductance $\kappa_0$ at room temperature for the central device, assuming transmission probability $\Theta(\omega)=1$ in a Landauer formula, given by
\be
\kappa_0=\int_0^{\omega_D} d\omega\;\left.\omega \frac{\partial b(\omega)}{\partial T}\right|_{T=T_{room}}.
\label{kappao}
\ee 
Here the left lead is set at $T_L=T_{room}+\Delta T$, the right lead at $T_L=T_{room}$ and $\Delta T\to 0$. 

As mentioned in the Introduction, we can not treat thermal conductance as a constant for large $\Delta T$. Instead, we consider the non-linear analog given by
\be
\kappa_{nl}=\frac{J_{net}}{\Delta T}; \;\;\; J_{net}\equiv \int_0^{\omega_D} J(\omega) \;d\omega.
\label{kappanl}
\ee
The total current $J_{net}$ is a function of disorder as well as the lead temperatures. Therefore the non-linear thermal conductance will also be a function of not only the disorder parameters of the wire, but also the thermodynamic parameters of the leads.

\subsection{Frequency dependent current}

The most important effect of the localized phonon is to provide a frequency-dependent scattering rate, as shown (for zero temperature) in [\onlinecite{ma}]. Here we consider the frequency dependent current for smooth vs rough nanowires, for different temperatures of the leads.


For simplicity, we assume the phonon-phonon coupling parameter to be given by $W_0=1$. Because of the universality of the disorder parameter as observed in [\onlinecite{mm}], the surface roughness can be characterized by a single dimensionless parameter 
\be
\zeta\equiv \frac{\sqrt{l_c d}}{h}
\label{xi}
\ee
where $h$ is the mean corrugation height, $l_c$ is the correlation length and $d$ is the diameter of the wire. Clearly larger $h/d$ and smaller $l_c$ implies higher disorder, so smaller vales of $\zeta$ corresponds to larger strengths of surface disorder. Although we fixed our diameter, we have kept the diameter dependence in (\ref{xi}) to emphasize that the effects of changing the diameter can be incorporated in this context within this framework by including it in the disorder parameter.

In our model, the width of the localized phonon $\Gamma$ is an independent phenomenological parameter. However since it arises from the mapping of the surface roughness, it also reflects the universality in terms of disorder, as shown in [\onlinecite{mm}], where it was argued that $\Gamma\propto l_c^{1/2}d^{3/2}/h=\zeta d$. We have therefore chosen the disorder dependence of $\Gamma$ by using $\Gamma=\Gamma_0\zeta d$ where $\Gamma_0$ depends on the coupling to the lead. In other words, $\Gamma_0$ is an independent phenomenological parameter that determines how strongly the localized phonon couples to the leads, but the dependence of $\Gamma$ on surface disorder is chosen to reflect the fact that larger disorder (smaller $\zeta$) should lead to a smaller width $\Gamma$. Note that we have a fixed diameter, and for all our calculations we choose $d=64$ nm  as a typical diameter in experiments on ELE nanowires, and $\Gamma_0=10^{-4}$.

\begin{figure}
\begin{center}
\includegraphics[width=0.48\textwidth]{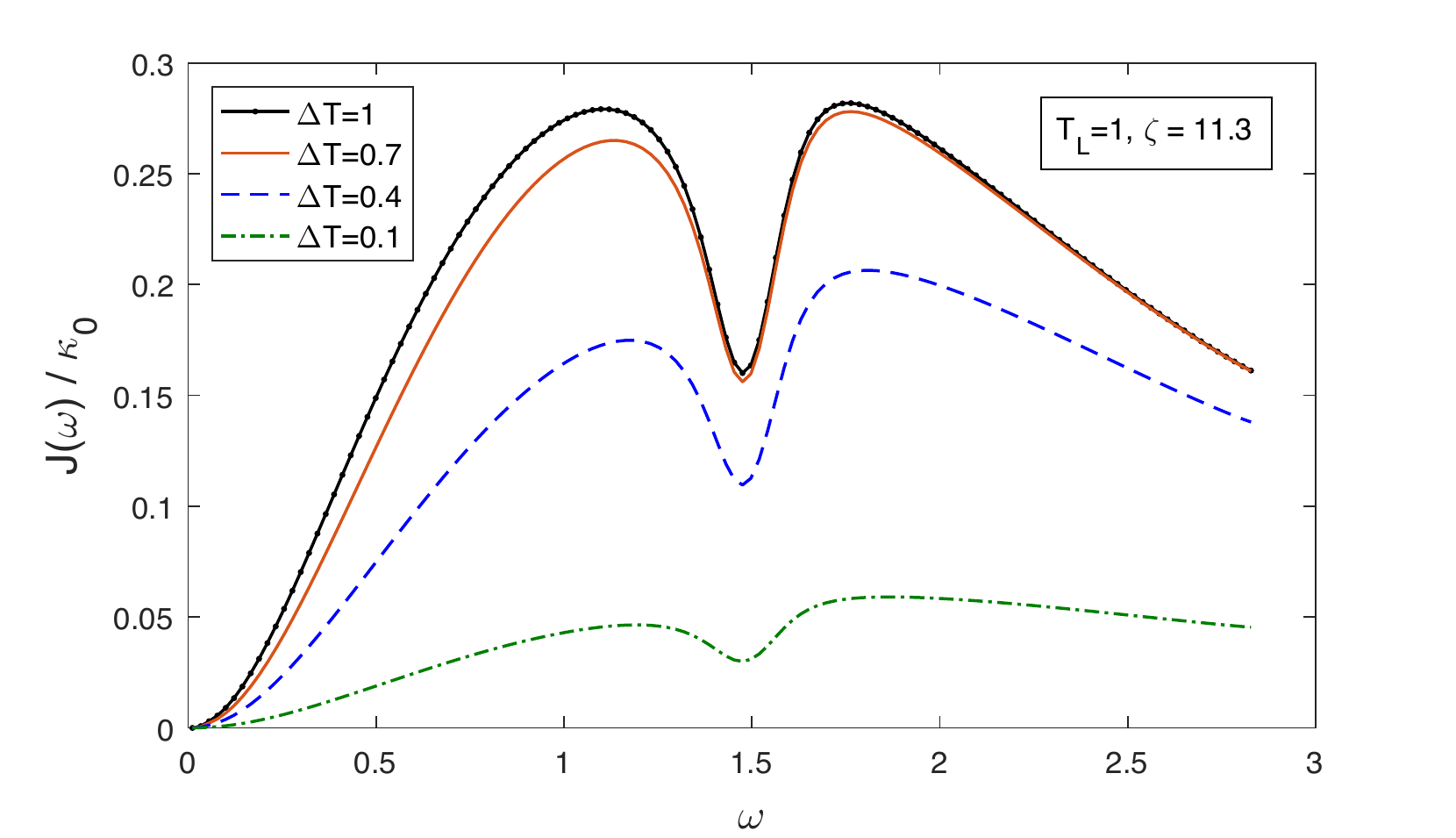}
\includegraphics[width=0.48\textwidth]{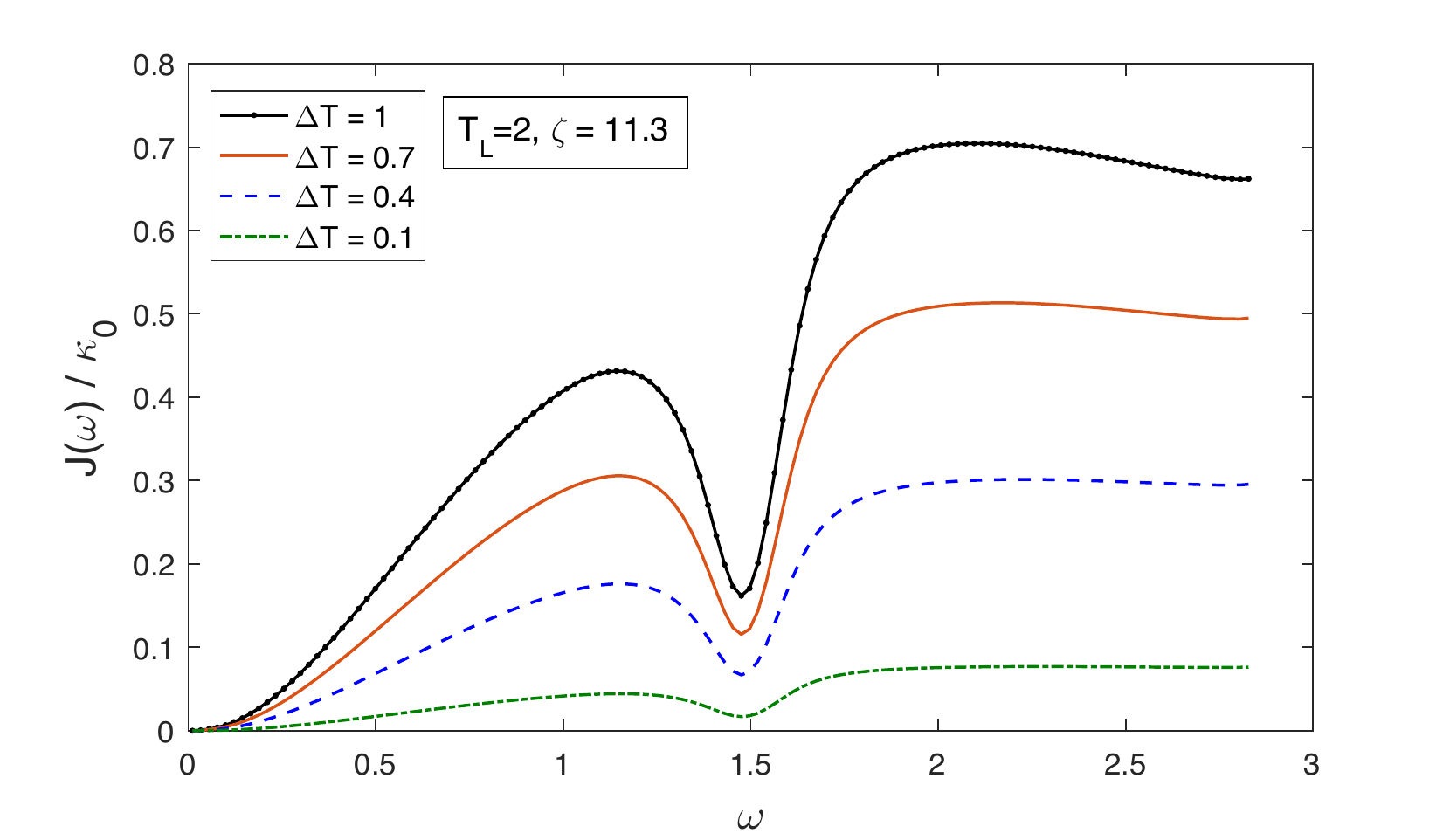}
\end{center}
\caption{
(Color online) Normalized thermal current $J(\omega)/\kappa_0$ as a function of frequency $\omega$ for a wire with surface roughness $\zeta=11.3$, for various temperature differences $\Delta T$. Temperature of the left lead $T_L=1$ in the top panel and $T_L=2$ in the bottom panel. }
\label{DeltaT2}
\end{figure}

In Figure \ref{DeltaT2}  we show the effects of changing $\Delta T$ for two different values of $T_L$ on the frequency dependent current. 
For definiteness, in this plot we choose $\zeta=11.3$ as a fixed disorder  which can be achieved in real experiments. As noted in [\onlinecite{mm}],  experiments on silicon nanowires with ELE surface disorder \cite{hochbaum,lim} have one sample with $h=2.3$, $l_c=8.9$ and $d=77.5$ in units of nm, corresponding to $\zeta=11.3$. 

The qualitative features of the overall frequency dependence of $J(\omega)$ is easy to understand. The low-frequency limit can be obtained analytically with Im $\Sigma^r\propto \omega^2$ and $D^{\lessgtr}\propto \omega$, leading to the current $J(\omega)\propto \omega^3$. The high frequency limit (where the frequency is large compared to the temperature of the hot lead) is dominated by the Bose distribution functions. The thermal current increases with increasing $\Delta T$ as expected, with maximum around $\omega\sim T_L$. 

The scattering from a localized phonon produces an expected dip at the localized phonon frequency $\Omega$. Clearly, if we included many localized phonons, there would be many dips corresponding to those frequencies. In principle, there should be a distribution of frequencies and one should integrate over that distribution. In the absence of a microscopic model of the surface roughness, such details can not be included in our calculation. On the other hand, our result suggests that  experimental observation of such identifiable or overlapping dips would provide a better understanding of the distribution of the localized phonons associated with the surface roughness in the system. Note that the dips due to the scattering from a localized phonon become more pronounced with increasing $\Delta T$. Thus experiments done at various large values of $\Delta T$ would have a better chance of observing the effect.

\begin{figure}
\begin{center}
\includegraphics[width=0.48\textwidth]{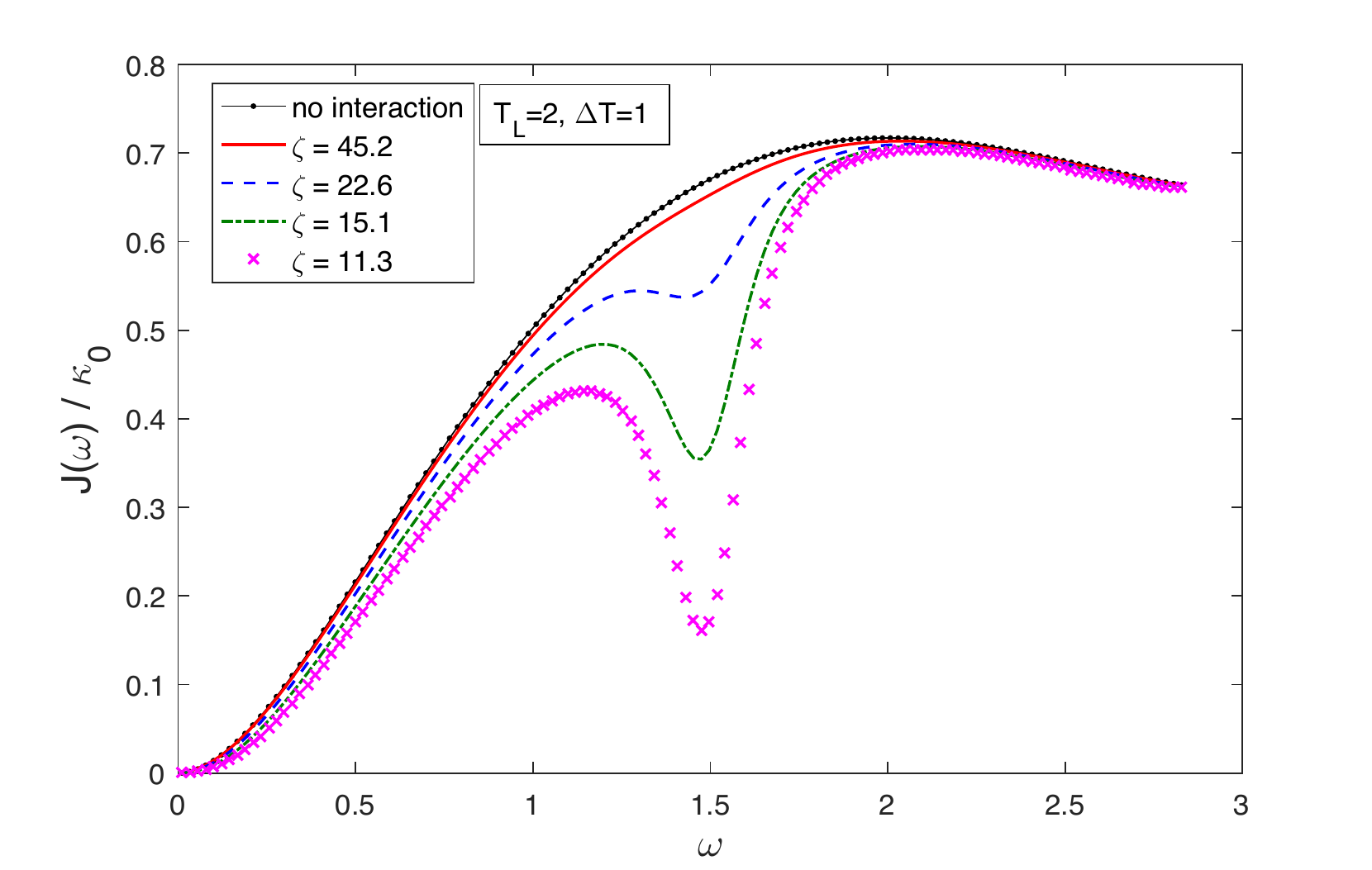}
\end{center}
\caption{
(Color online) Normalized thermal current $J(\omega)/\kappa_0$ as a function of frequency $\omega$ for a wire with fixed $T_L=2$ and $\Delta T=1$, for various surface roughness parameters $\zeta$.}
\label{Deltazeta}
\end{figure}

Finally, we show the effects of changing the strength of surface disorder for a fixed set of temperatures in the leads. 
Figure \ref{Deltazeta} shows the thermal current for fixed temperatures of the two leads as the disorder parameter is changed. We choose $T_L=1$ and $\Delta T=1$ for comparison with previous plots.
Note that smaller $\zeta$ corresponds to larger disorder.

\subsection{Non-linear thermal conductance}

\begin{figure}
\begin{center}
\includegraphics[width=0.48\textwidth]{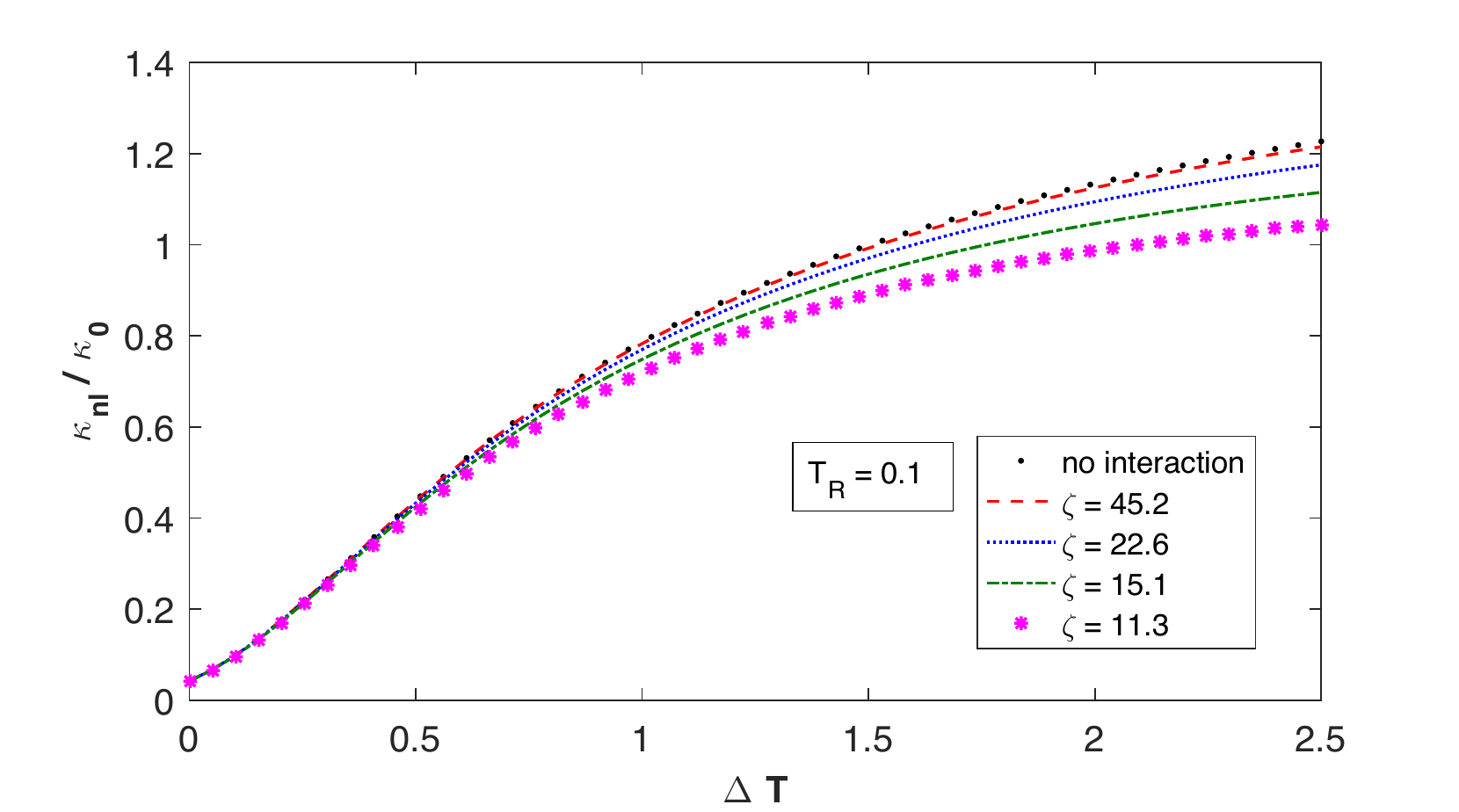}
\end{center}
\caption{
(Color online) Normalized non-linear thermal conductance $\kappa_{nl}/\kappa_0$ as a function $\Delta T$ with fixed right (cold) lead temperature $T_R=0.1$ for various surface roughness parameters $\zeta$.}
\label{Jnet-nonlinear}
\end{figure}

\begin{figure}
\begin{center}
\includegraphics[width=0.48\textwidth]{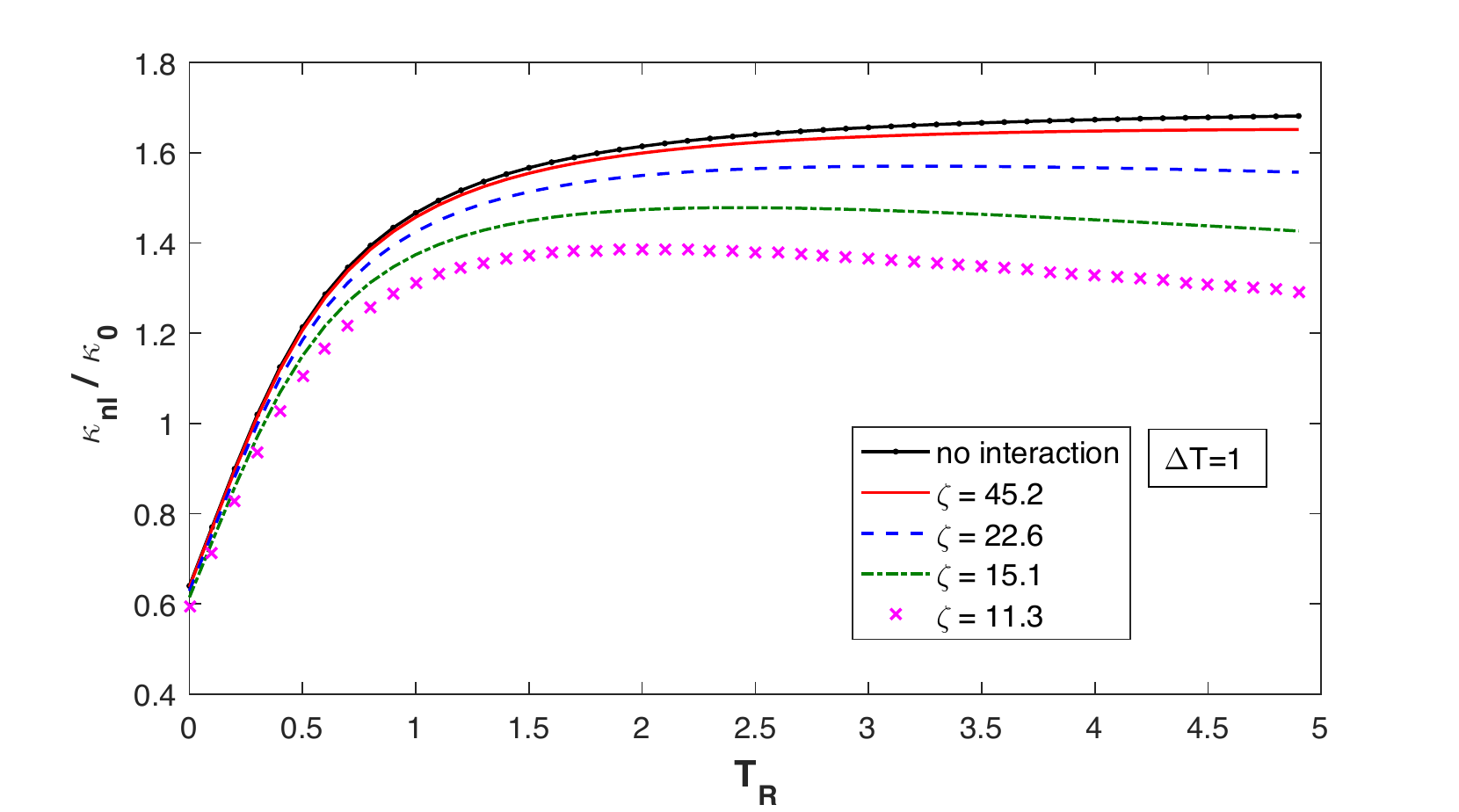}
\end{center}
\caption{
(Color online) Normalized thermal conductance $\kappa_{nl}/\kappa_0$ as a function of the right (cold) lead temperature $T_R$ for a fixed $\Delta T=1$, for various surface roughness parameters $\zeta$.}
\label{Jnet1}
\end{figure}

The linear response theory assumes that the thermal conductance is a constant, determined entirely by the  properties of the material. We find that the general  non-linear thermal conductance $\kappa_{nl}$ as defined in Eq.~(\ref{kappanl}) is a non-trivial function of the lead temperatures as well.  In Figure \ref{Jnet-nonlinear} we plot $\kappa_{nl}$ as a function of $\Delta T$. It clearly shows the non-linearity over a wide range of $\Delta T$ which depends on the strength of disorder and  emphasizes the fact that a linear response theory is not adequate. 
Clearly the non-linear regime should be experimentally accessible, and practically important.
Figure \ref{Jnet1} shows the effect of disorder on $\kappa_{nl}$ as a function of the lead temperature $T_R$ for a fixed value of $\Delta T=1$. 
The non-interacting (smooth wire) current keeps increasing at large temperature, while it either saturates or decreases with increasing temperature for rough surfaces.  The decrease is larger for larger surface disorder. As shown in [\onlinecite{mm}], even with surface disorder present, a simple Landauer-like formula in the linear-response regime leads to a saturation of the thermal conductivity at temperatures much larger than the localized phonon frequency. The eventual decrease must therefore arise from the non-linear regime where the dip is much larger at the localized phonon frequency, as seen in Figure \ref{DeltaT2}.

\begin{figure}
\begin{center}
\includegraphics[width=0.48\textwidth]{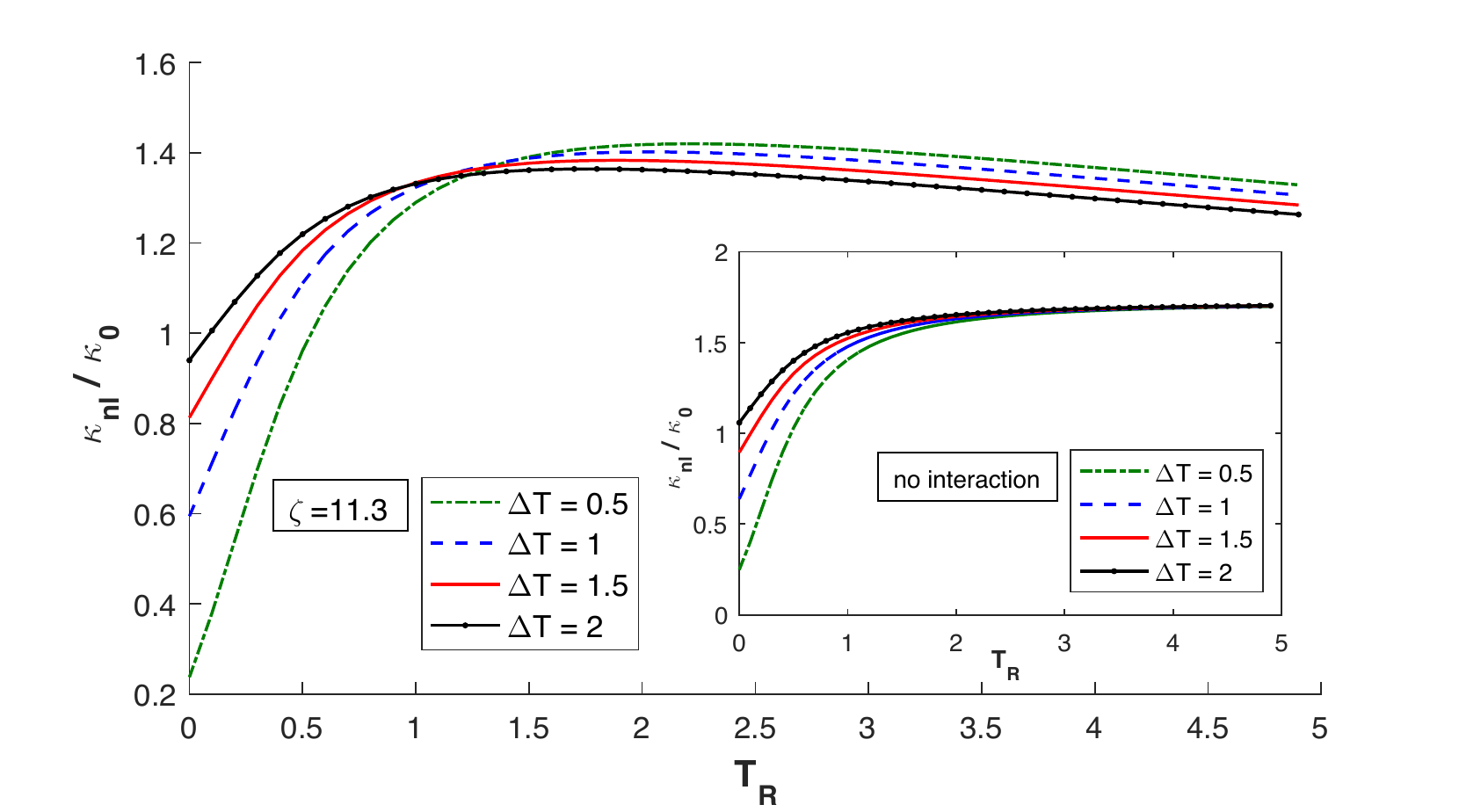}
\end{center}
\caption{
(Color online) Normalized non-linear thermal conductance $\kappa_{nl}/\kappa_0$ as a function of the right (cold) lead temperature $T_R$ for a fixed surface roughness parameter $\zeta=11.3$, for various values of  $\Delta T$. Inset shows the same for a non-interacting (smooth) wire.} 
\label{Jnet2}
\end{figure}

Figure \ref{Jnet2} shows $\kappa_{nl}(T)$ as a function of the lead temperature $T_R$ for various $\Delta T$, for a fixed value of the surface roughness parameter $\zeta$. 
The difference in the resulting curves show the importance of non-linearity in the thermal conduction and the need for a fully non-equilibrium calculation that goes beyond the Landauer approach.  

It is difficult to compare our results for the highly non-linear thermal conductance with existing experiments which are typically done only in the limit $\Delta T\to 0$. One trend that is clear from experiments is that even in the linear response regime the thermal conductivity in wires without surface disorder (the VLS wires \cite{huang}) keep increasing with increasing temperature, while it either saturates or decreases in wires with surface disorder (ELE \cite{hochbaum} and EBL \cite{huang} wires) at the same temperaure, consistent with Figure \ref{Jnet1}. The difference from the non-interacting case is larger for larger surface disorder, and the results should be qualitatively valid for a more realistic model with a distribution of localized frequencies. 
Experiments are not currently available for thermal current at large finite temperature differences for  nanowires with rough surfaces. Since the effects of scattering from localized phonons are more pronounced at larger $\Delta T$, experiments done at larger $\Delta T$ will have a better chance of observing the effects of surface disorder.


\section{Summary and discussion}

We use a previously worked out mapping where the effects of scattering of phonons from a nanowire with a rough surface can be considered as equivalent to scattering from randomly situated localized phonons in the bulk of a smooth nanowire. We use the resulting model of interaction  between propagating and localized phonons proposed in [\onlinecite{ma}] to evaluate the thermal current  between two ideal semi-infinite leads connecting a nanowire in the presence of arbitrary temperature differences between the leads at various surface disorders. We use the non-equilibrium Green's function techniques which  allow us to go beyond the linear response regime. The lesser and greater Green's functions for the propagating phonons are calculated self consistently, so that an infinite summation of the leading non-crossing diagram is included.

We limit ourselves to scattering from a single localized phonon to illustrate the qualitative effects. The frequency dependence of the thermal current contains clear signature of scattering from the localized phonon. We propose that comparing our results with experiments should lead to an understanding of the distribution of such localized phonons in nanowires with rough surfaces, where the roughness can be characterized by a single disorder parameter $\zeta$. Such experiments are not yet available. The effects are more pronounced for larger temperature differences between the two leads, so we suggest that experiments done in this non-linear regime have a better possibility of observing them. 
We find that $\kappa_{nl}=J_{net}/\Delta T$ is a highly non-linear function of $\Delta T$.  The change in this function with (cold) lead temperature or with the disorder parameter should be experimentally observable.    Our results explain the difference in the temperature dependence in experiments on nanowires with smooth vs rough surfaces, the so called VLS vs ELE or EBL wires.  The results also show the importance of going beyond the linear response regime in the context of  thermoelectric devices. Systematic experimental studies of the thermal current with different values of $\Delta T$ as well as different strengths of surface disorder characterized by $\zeta$ will be very valuable in understanding the role of localized phonons in the context of thermoelectricity.

In this work we are primarily interested in a qualitative understanding of the effects of surface disorder on thermal transport. For a more quantitative understanding, there are several areas where the present calculation can be improved. For example, in our calculations we only focus on the contribution of interactions to the imaginary part of the self energy, neglecting all corrections to the real part. This might be a problem if the corrections are large enough e.g. to move the localized phonon out of the propagating band. In our phenomenological model, any such shift is included in the choice of the parameters. However,  with future inputs from experiments, more realistic values of the parameters for the localized phonons can be used where corrections to the real part would be important. In addition, our calculations include only one localized phonon. Although it already  captures the important qualitative features, inputs from experiments can suggest possible distribution of frequencies and/or the widths which can be incorporated in the model. Finally, we have not done a fully two-dimensional calculation in order to avoid computational complexity. We believe that this is not crucial for the present model where motion perpendicular to the length is included in the mapping that takes care of the scattering from the surface. However, it restricts us from obtaining the diameter dependence of the thermal current. It should be possible to consider a more realistic model of a nanowire and extend the present calculation to address the above limitations.

\section{acknowledgment}

SA is grateful to S. Hershfield and A. Hebard for valuable suggestions. KAM acknowledges useful discussions with P. Marko\v{s}.

\end{document}